# Research evaluation with ChatGPT: Is it age, country, length, or field biased?

Mike Thelwall, Zeyneb Kurt: Information School, University of Sheffield, UK.

Some research now suggests that ChatGPT can estimate the quality of journal articles from their titles and abstracts. This has created the possibility to use ChatGPT quality scores, perhaps alongside citation-based formulae, to support peer review for research evaluation. Nevertheless, ChatGPT's internal processes are effectively opaque, despite it writing a report to support its scores, and its biases are unknown. This article investigates whether publication date and field are biasing factors. Based on submitting a monodisciplinary journal-balanced set of 117,650 articles from 26 fields published in the years 2003, 2008, 2013, 2018 and 2023 to ChatGPT 4o-mini, the results show that average scores increased over time, and this was not due to author nationality or title and abstract length changes. The results also varied substantially between fields, and first author countries. In addition, articles with longer abstracts tended to receive higher scores, but plausibly due to such articles tending to be better rather than due to ChatGPT analysing more text. Thus, for the most accurate research quality evaluation results from ChatGPT, it is important to normalise ChatGPT scores for field and year and check for anomalies caused by sets of articles with short abstracts.

## Introduction

Expert review of the quality of published research outputs is time-consuming and seems to have become increasingly prevalent as part of the procedures for appointments, promotion, and tenure. In some countries, including the UK, post publication expert review also occurs systematically in national research evaluation exercises (e.g., Technopolis, 2024). Nevertheless, bibliometric indicators are sometimes used instead of expert review because it is impractical to read work on a large enough scale, especially if the purpose is exploratory or if the review costs would outweigh the value of the results (e.g., for comparing national performance to competitor countries: Aksnes et al., 2012, Gov.UK, 2022). Between these two extremes, the use of bibliometrics to support expert review is consistent with the Leiden Manifesto (Hicks et al., 2015), the Metric Tide (Wilsdon et al., 2015), and by the Coalition for Advancing Research Assessment (coara.eu). This can improve score accuracy when experts are uncertain and might be used to reduce the burden on reviewers (e.g., by using two instead of three) when the bibliometrics are known to be accurate enough. Improving the value of bibliometric indicators, the focus of much scientometric research, or finding an improved alternative, as in the current paper, therefore has the potential to save expert time and/or increase the accuracy of their decisions.

Previous research has found that ChatGPT might provide an alternative to citation-based indicators in support of human expert review, or for some theoretical investigations of science that currently use citation analysis (e.g., Chen et al., 2015). This is because, when fed with expert review instructions, ChatGPT can provide quality scores for journal articles that positively correlate with human expert scores in all fields of scholarship except clinical medicine (Saad et al., 2024; Thelwall, 2024ab; Thelwall & Yaghi, 2024). Moreover, ChatGPT's scores have a higher correlation with human expert scores than do citation-based indicators or traditional Artificial Intelligence (AI) in most fields (Thelwall & Yaghi, 2024). Despite these promising statistical results, and the fact that AI evaluations of academic proposals are now supporting at least one funding agency (Carbonell Cortés, C. et al., 2024), nothing is known about any biases in ChatGPT results. In other contexts, the potential for AI systems to replicate human prejudices and even invent new partialities has been shown (Ntoutsi et al., 2020). It would therefore be unwise to assume that ChatGPT's research quality evaluations are exempt or to use it for important evaluations without prior investigations into potential biases.

In response to the above concern, this article primarily focuses on one important potential bias, publication year. When human experts assess articles for research quality, they can be expected to consider the publication date when evaluating novelty and significance because, for example, earlier articles on a topic would tend to be more important than later articles, other factors being equal. They might be judged more original by covering an under-researched or new topic or more influential by making first investigations that later contributions then refined. This article therefore assesses whether ChatGPT's average quality scores vary based on the publication years of articles, which ChatGPT is typically not told. The main concern is that it would tend to give lower scores to older articles because it does not consider their value relative to when they were published. Field differences in citation rates are well known in citation analysis (Dunaiski et al., 2019) and so this is another source of potential bias that should be investigated. Finally, since optimal ChatGPT quality scores are obtained from the title and abstract of an article, ignoring its full text, it is logical to also assess whether the combined length of these associates in any way with ChatGPT scores. The main research questions are therefore as follows.

- RQ1: Do average ChatGPT scores vary systematically by year?
- RQ2: Do average ChatGPT scores vary between fields?
- RQ3: Do average ChatGPT scores associate with title and abstract length?

Whilst country differences may not be due to biases but instead a result of national factors, such as the degree of financial support for research, it is important to identify any such differences for text-based indicators because of the possibility that countries are disadvantaged if they have few native English speakers. If English speaking countries consistently outscored others with ChatGPT, then this would suggest a language bias.

- RQ4: Do average ChatGPT scores vary between countries?

This article also reports whether ChatGPT scores associate with citation counts because the biases of citation-based indicators have been extensively investigated (e.g., Didegah, 2014) and this evidence provides additional context to the main results.

- RQ5: Do ChatGPT scores correlate with citation counts?

## Methods

The research design was to take samples of research articles from a range of fields and a range of widely spaced years, controlling for as many variables as possible, then compare the average scores by year and field for an initial descriptive analysis and then use regression to compare the effects of all variables simultaneously. A simultaneous comparison of all variables through regression is essential to guard against one bias being a second order effect of another.

Although previous studies have shown that more accurate scores can be obtained by ChatGPT if articles are submitted multiple times (Thelwall, 2024ab), accuracy is not of primary interest for the current study (except RQ5, which is provided for context) and so each article was submitted only once.

*Data*

A dataset was needed of articles published in different years but with the same average quality each year. Unfortunately, it is impossible to be sure that the average quality of two sets of papers published at different points in time is the same because all aspects of academic scholarship evolve. The simplest approach is to take a random sample from the subject categories of a scholarly database, such as the Web of Science (WoS) or Scopus, that are quality controlled. This is not ideal, however, because the coverage of these databases can shift substantially when a new tranche of journals is added (Moed et al., 2018). The situation is exacerbated if indexing policy changes occur, such as to add journals from an emerging research country. A better, but still imperfect, approach might be to select one or a small set of core journals from each field to take samples from, on the basis that the quality of journals seems to be relatively stable. This is unproven and journals can change in nature when a new editor takes over or perhaps after a deliberate shift in scope (Martin, 2016; Wilhite et al., 2019) but there does not seem to be a more reliable approach.

Based on the above discussion, a journal-based sampling method was designed to minimise likely biases. Scopus was chosen as the database to select the data from, although WoS would have been equally appropriate. The years 2003, 2008, 2013, 2018 and 2023 were selected with five-year gaps to identify overall trends over the past two decades. A relatively long period was needed in case changes between adjacent years were too small to be detected statistically. The Scopus records were downloaded in January-March 2024. The 26 non-general categories in Scopus were used as the fields to analyse. Field definitions are controversial (e.g., Waltman & Van Eck, 2012) and fields evolve over time (Porter & Rafols, 2009), but the Scopus definitions are at least transparent in the sense that the underlying sets of journals are openly published (www.elsevier.com/en-gb/products/scopus/content). Each field was represented from the set journals that had articles published in each of the five years. Journals were selected when they were not also classified in other Scopus categories to avoid overlaps between fields. Although some overlap would be reasonable, this rule has the advantage of excluding huge multidisciplinary journals that would have many articles that are out of scope for some of the fields in which they are classified.

Article titles and abstracts were extracted from Scopus and automatically cleaned of copyright statements using Webometric Analyst (github.com/MikeThelwall/Webometric_Analyst) for submission to ChatGPT. Some articles in Scopus are short form, letters or corrections and are not flagged as such in the database. These typically seem to have a short abstract. For this reason, and because some articles did not have an abstract in Scopus, a minimum abstract length threshold of 786 characters, after removing abstract copyright statements, was set for articles to be selected. For the journals selected, this represented ignoring 25% of Scopus articles with short or no abstracts. This threshold was chosen as a round number (in percentage terms) and as, heuristically, long enough to judge the content of a full article. For example, a threshold of 10% equates to 403 characters which is short, and 20% equates to 682 characters, which still seemed too short for a meaningful description of an academic article.

Articles were selected at random from all journals matching these criteria, for a total of 1000 for each year. The random selection process was conducted by journal, so each journal had the same number of articles in each year. Because of this, the overall number of articles per broad field varied slightly from the target 1000.

In summary, for each of the years 2003, 2008, 2013, 2018 and 2023 and each of the 26 non-general Scopus categories, up to approximately 1000 articles were selected with a random number generator (without replacement) from the set of all articles with substantial abstracts in journals that were exclusive to that category and that had published in all five years (Figure 1). This gave a total of just under 1000x5x26 articles to submit to ChatGPT.

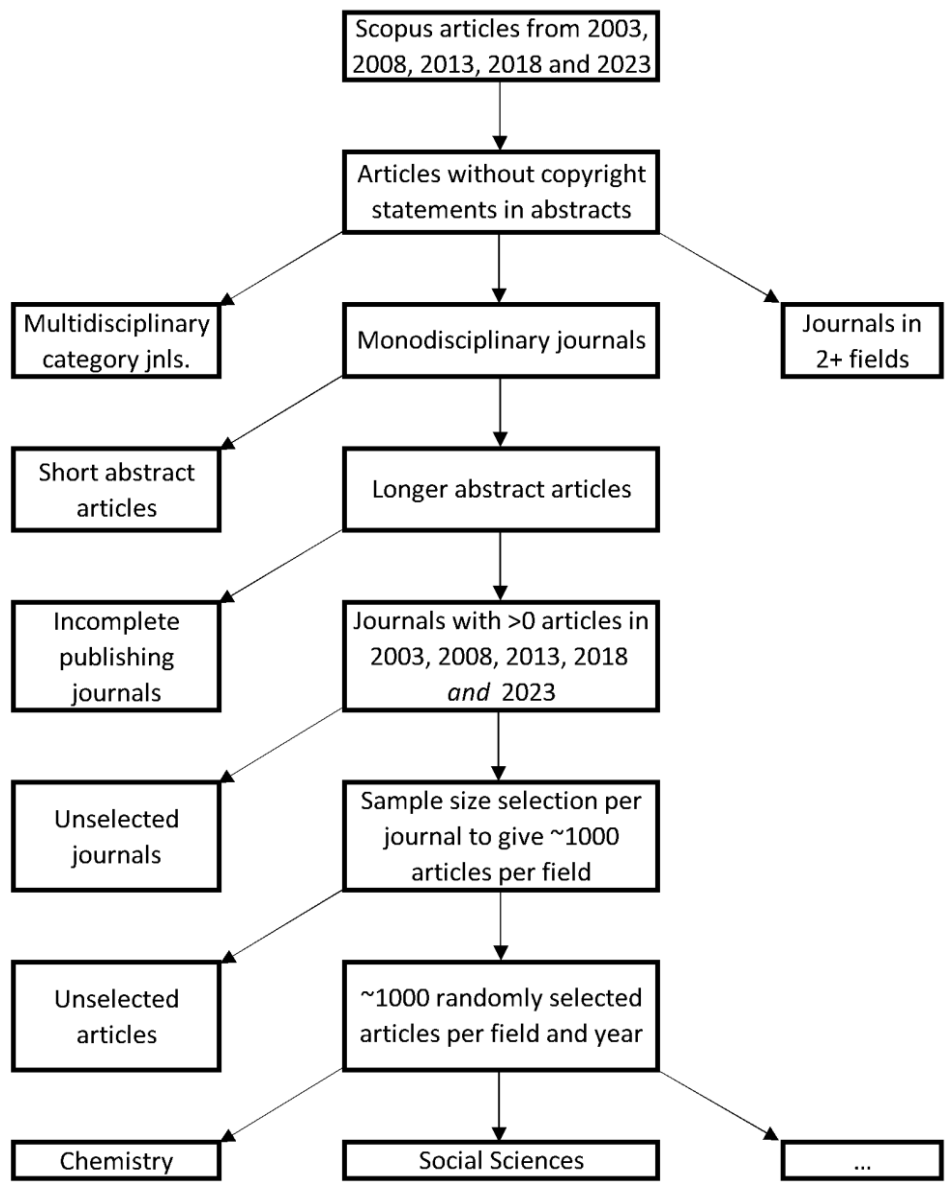

Figure 1. The main sample selection stages.

## ChatGPT procedure

The ChatGPT setup was the same as for a previous analysis of journal articles submitted to the UK REF (Thelwall & Yaghi, 2024). In summary, ChatGPT was supplied with the appropriate evaluation guidelines given to REF assessors. These guidelines were used because they have been previously tested with ChatGPT and are reasonably detailed guidelines leading to specific scores (1*, 2*, 3* or 4*). There are four guidelines, and the most relevant one was used for each field. In particular, the health and life sciences guidelines (REF Panel A) were applied to Scopus fields 11, 13, 24, 27, 28, 29, 30, 32, 34, 35, 36; the physical sciences and engineering guidelines (REF Panel B) were applied to fields 15, 16, 17, 18, 19, 21, 22, 23, 25, 26, 31; the social sciences guidelines (REF Panel C) were applied to 14, 20, 33; and the arts and humanities guidelines (Main Panel D) were applied to 12. These numbers can be converted to fields by adding 00 and consulting https://service.elsevier.com/app/answers/detail/a_id/15181/supporthub/scopus/.

Each article was submitted to ChatGPT 4o-mini using the ChatGPT API. This interface does not use the submitted material for model building (openai.com/consumer-privacy), which avoids any potential feedback within the experiment as well as any copyright concerns. The scores were extracted by a program designed to apply a set of pattern-matching rules to identify scores from ChatGPT reports (github.com/MikeThelwall/Webometric_Analyst, AI menu). When its rules failed, the first author was shown the ChatGPT output and prompted for a number. The scores were usually whole numbers, but occasionally separate scores were provided for the three individual aspects of research quality considered in the REF (originality, rigour, significance) and these where then averaged for the overall score.

### Descriptive analysis

Average ChatGPT scores were calculated for each year/field combination. The results were then graphed for each field to identify any changes over time and between fields. Pearson correlations for ChatGPT score and the log of the abstract length were then calculated for each year and field combination to identify any large scale underlying trends.

### Regression

Ordinary least squares regression was used to check whether publication year could be a second order effect of changes over time in abstract length or author affiliation country, and to check for the importance of the latter variables. For this, ChatGPT score was used as the dependant variable, with the following independent variables: publication year; title and abstract length combined (number of characters); first author country (ten binary-coded variables). First author country was recorded for the ten countries with the most Scopus publications. Although it would have been possible to include a separate binary variable for each country in the world, this would lead to overfitting so a focus on the top ten publishing countries seems adequate to capture the main national influences.

Both abstract and title length and the log of abstract and title length were tried as independent variables. They had high Variance inflation factors (VIFs) between them (typically above 20) so only one could be retained due to the multicollinearity. The log version was used because it was statistically significant in more regressions combining both. After this decision, VIFs were calculated for all variables, with most values being close to 1 and none above 2, so the multicollinearity is not strong enough to correct for. Q-Q plots were broadly consistent with normally distributed residuals.

## Results

### RQ1-3: Descriptive analysis

The descriptive analysis of average ChatGPT scores shows differences between fields and systematic differences between years. For all 26 broad fields, the ChatGPT average for 2023 was higher than for 2003, and in 101 out of 104 cases, the ChatGPT average for each year (2008, 2013, 2018, 2023) was higher than the average from the previous year. There was also a positive relationship between publication year and ChatGPT score at the article level, with Pearson correlation coefficients ranging between 0.077 (Mathematics) and 0.306 (Business, Management and Accounting), with a mean of 0.181. From the perspective of the article-level influence of publication year, the $R^2$ values corresponding to these correlation coefficients (not the linear regressions, which are discussed in the next section) show that the amount of variance in the ChatGPT scores explained by the publication year varied between 0.6% and 9.4%, with an average of 3.6%. This indicates a relatively small influence.

There are also substantial differences in average ChatGPT scores between fields (Figure 2). For example, the averages for all five years of the top three fields (Biochemistry, Neuroscience, Mathematics) are higher than the averages of all fields in all years. Conversely, the averages for all five years of the lowest field, Veterinary, are lower than the averages for all years of all fields except the bottom four fields.

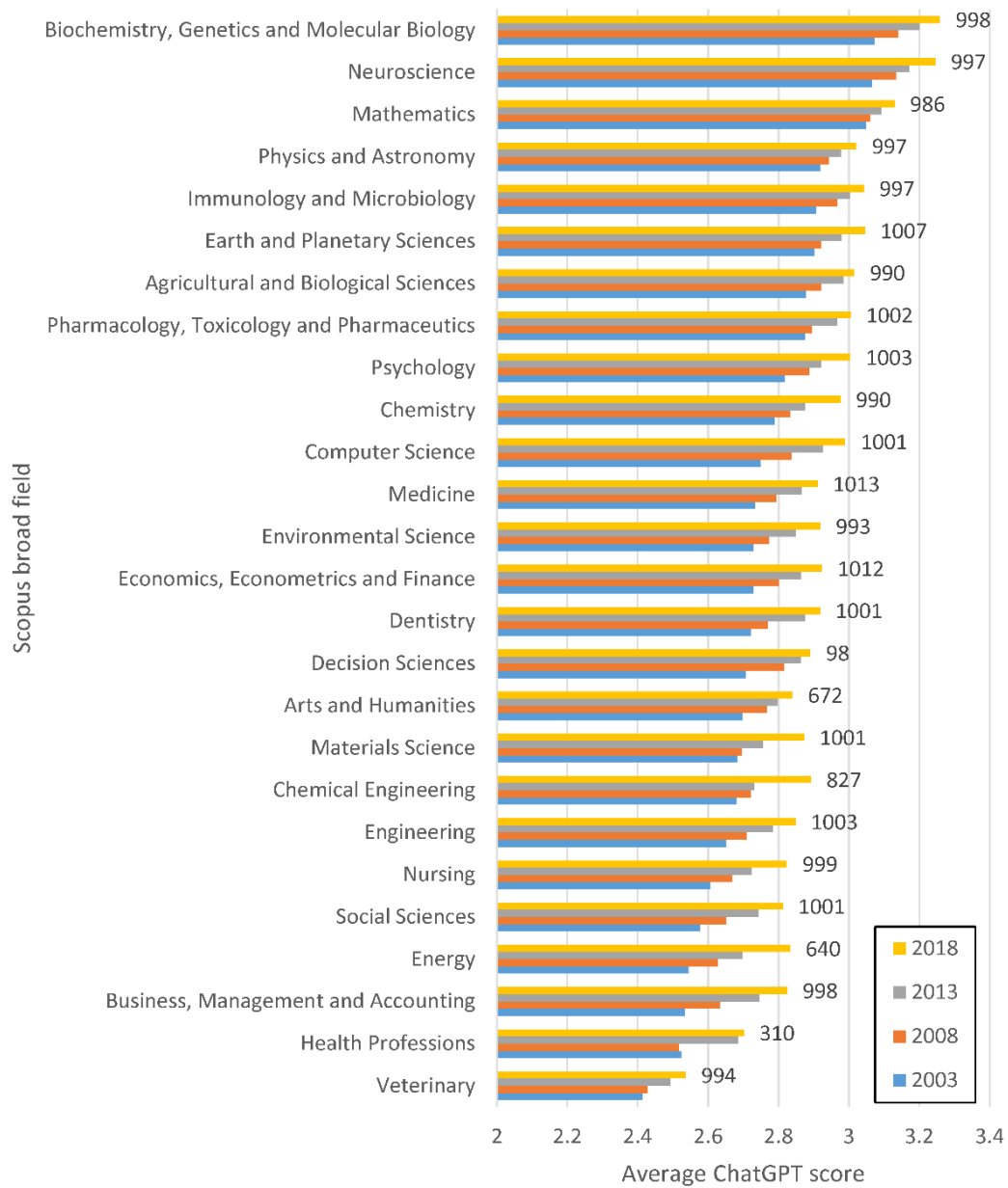

Figure 2. Average ChatGPT scores for articles from journals exclusively in each of 26 Scopus broad fields, by publication year. Fields are annotated with their sample sizes (number of articles in each of the five years).

Articles with longer abstracts tend to have moderately higher ChatGPT scores in some fields but not others, with a slight tendency for weaker correlations in more recent years (Figure 3). From the perspective of the article-level influence or association with abstract length, the $R^2$ values corresponding to these correlation coefficients (again, not the linear regressions) reveal that the amount of variance in the ChatGPT scores explained by abstract log-length variable varied between 0.0% and 11.7%, with an average of 3.3%. This indicates a smaller influence or association than for publication year overall, but larger in some fields.

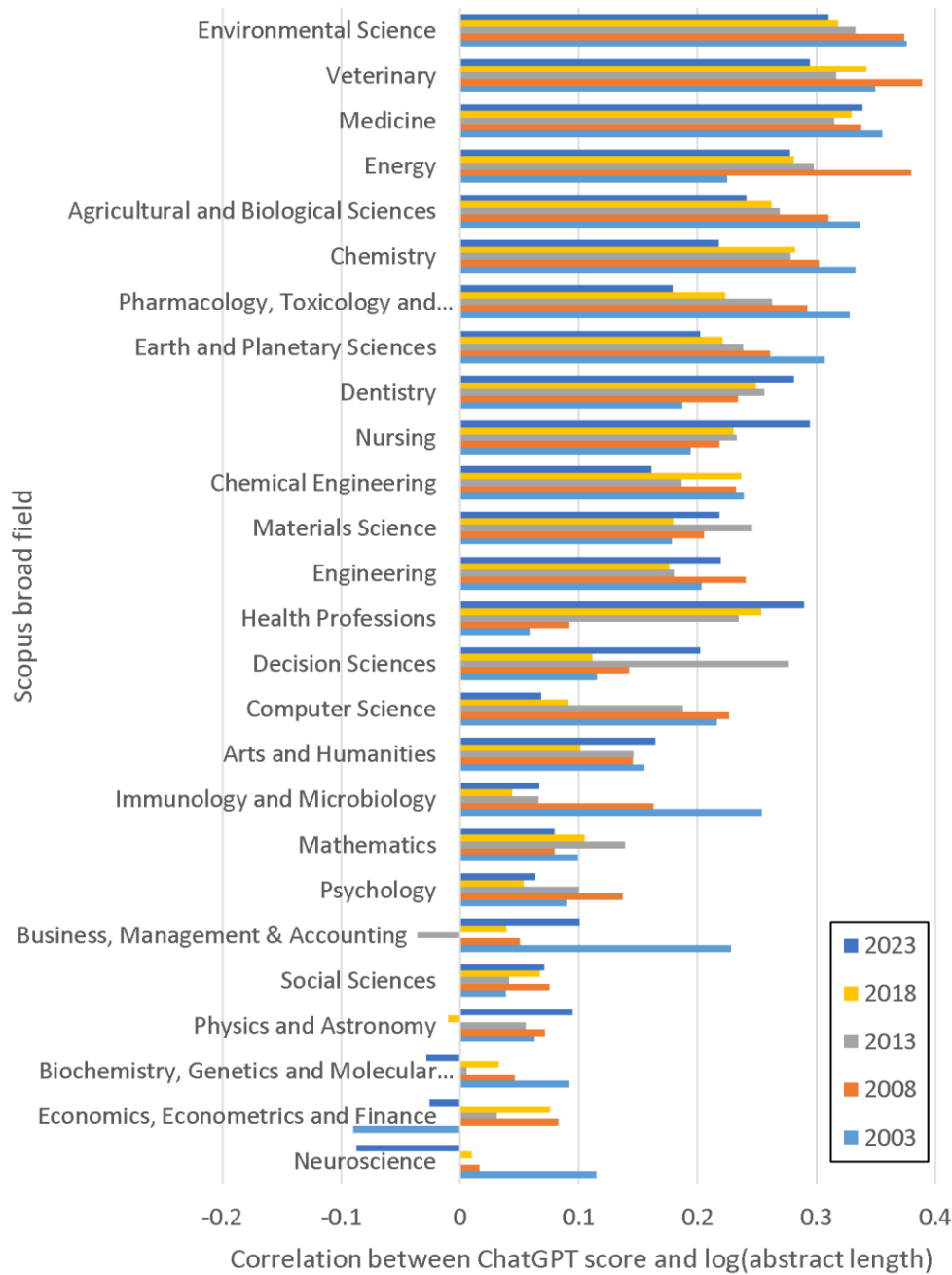

Figure 3. Pearson correlations between ChatGPT score and the log of the abstract length by year and field. Sample sizes are as in Figure 2.

## RQ1-4: Regression results

This section reports the aggregate results from the 26 separate field-based regressions. By taking all variables into account simultaneously, they help to exclude the possibility that any association is a second-order effect of one of the other independent variables.

In all 26 regressions, the year variable coefficient was positive and statistically significant, suggesting that the reason for the lower average ChatGPT scores for older articles is not due to changes in either the typical countries of the first authors or abstract length (Figure 4). Nevertheless, articles with longer abstracts tend to get higher scores in all 26 fields, and the coefficient is statistically significant in 23 of these fields. This is despite the minimum length abstract cutoff (see methods) that was designed to exclude short form contributions.

Although Canadian first authors were the only national set found to associate with higher ChatGPT scores in all fields (statistically significant in 21), most of the other countries with the most journal articles in Scopus also tended to be associated with higher ChatGPT scores. The main exception, India, has a low per capita research investment (Gross Expenditure on R&D as a percentage of GDP: https://data.uis.unesco.org/index.aspx?queryid=74), so its researchers are less supported. Considering the coefficients of Italy, France, and Germany, there is not a clear pattern of English-speaking nations having an advantage over others. Moreover, given that the REF criteria are important for UK academics, it is perhaps surprising that its authors do not dominate for the ChatGPT score derived from the same criteria.

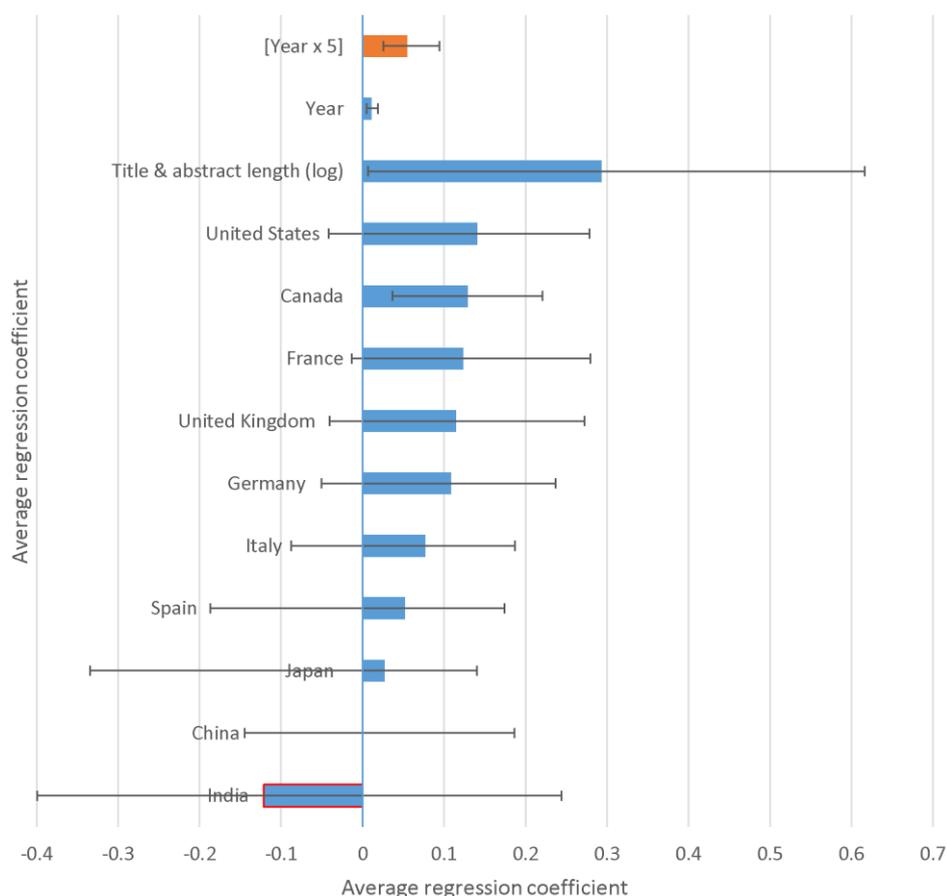

Figure 4. Summary of the coefficients extracted from 26 regressions for ChatGPT scores against the variables above (except Year x 5, added for illustrative purposes). Country names are for first authors, and only the ten countries with the most articles in Scopus were included.

Error bars show the minimum and maximum values from the 26 broad field estimates (not always statistically significantly different from 0). The coefficients can be compared between countries but none of the other coefficients should be compared.

The $R^2$ values of the models varied from 0.05 to 0.21, with a mean across fields of 0.12. Thus, on average, the factors considered above only account for 12% of the variation in the ChatGPT scores, so they do not have an overriding importance within a field. These values will be slight overestimates because cross-validation was not used to allow accuracy to be checked on different datasets to those used for model building.

## RQ5: Correlation with citation counts

For all fields, citation counts correlated positively with ChatGPT scores for the articles examined, although with substantial differences between fields (Figure 5). Correlation differences between years may be due to randomness inherent in the sample selection rather than due to systematic differences since the average of the 26 correlations for each year are similar without a monotonic increasing or decreasing trend (2003: 0.247; 2008: 0.232; 2013: 0.242; 2018: 0.233).

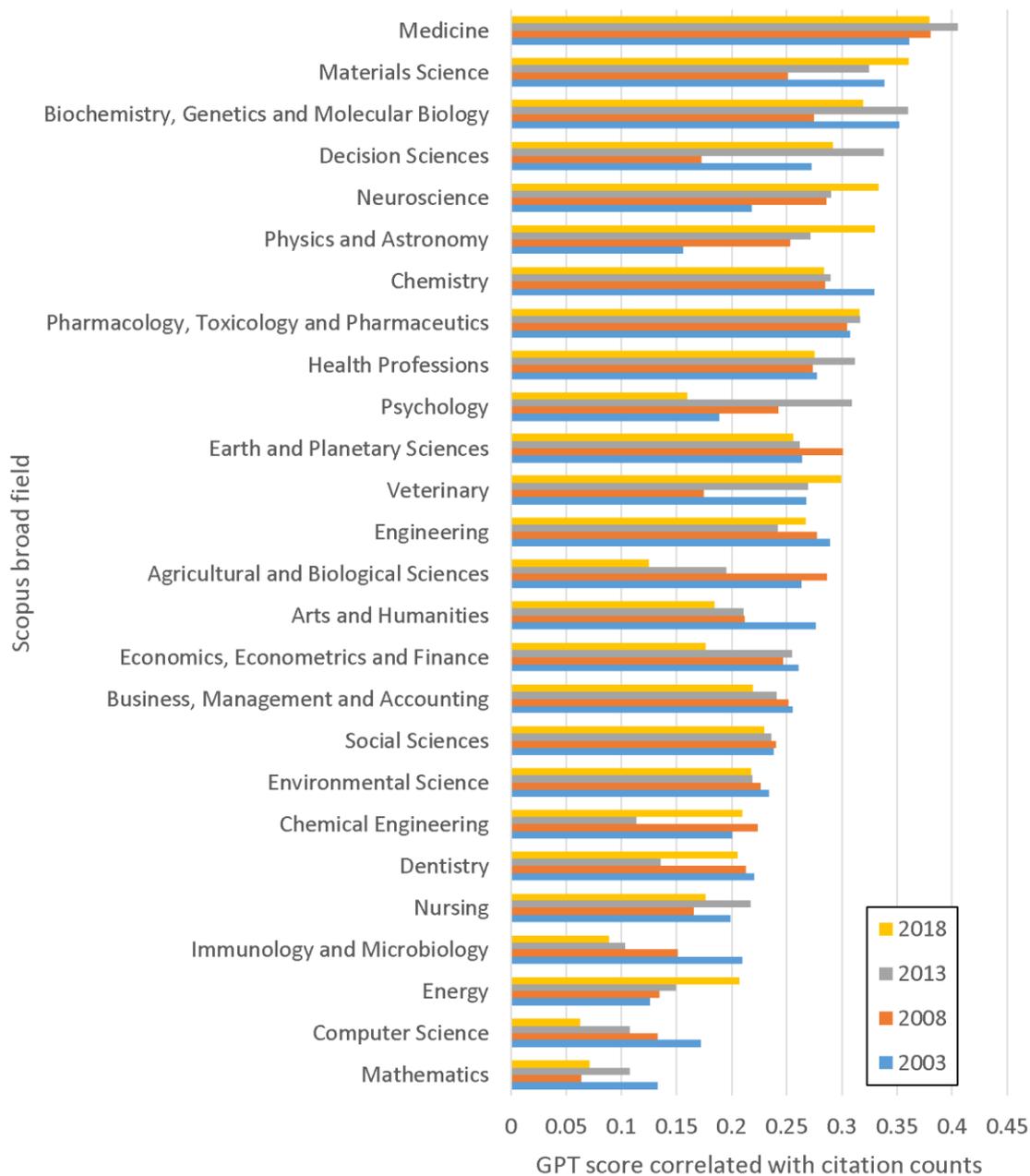

Figure 5. Spearman correlations between ChatGPT scores and Scopus citation counts. 2023 is excluded due to insufficient time for citations to accrue (Wang, 2013).

## Discussion

This article is limited by the sample selection method, which excludes articles in multidisciplinary journals and articles not indexed by Scopus. These may follow a different trend. The results may also be different for other LLM-based systems and may improve over time as each system evolves. The ChatGPT scores reflect the UK's national research quality definition but there are many others (Langfeldt et al., 2020), with Global South versions likely to be substantially different (Barrere, 2020). Moreover, if ChatGPT scores start to be widely used in research evaluation then authors may try to game them with their abstract designs, which may undermine the approach.

RQ1: The tendency for ChatGPT research quality scores to be higher for more recent articles has not been previously tested but the result is expected given that originality is a component of REF quality and is clearly time dependent. The converse relationship exists for recent citation counts, which accrue over time. Very old papers may get fewer citations, however, since articles mostly cite recent work and there were shorter reference lists in the past (Larivière et al., 2008).

RQ2: Field differences in average ChatGPT scores have not been found before but were to be expected given very different field norms in publishing styles (Puuska, 2014).

RQ3: In theory, the tendency for articles with longer abstracts to get higher ChatGPT scores could be a direct effect of better articles needing longer abstracts, a side effect of better journals allowing longer abstracts, or the inclusion of short form contributions with shorter abstracts in some or all fields in the dataset. An examination of the five fields with the highest correlations in Figure 3 suggests that the second and third options play a role as follows.

- **Better journals allow longer abstracts**: Journals publishing longer articles tend to attract higher ChatGPT scores in 2023 Environmental Science (average journal level Pearson correlation between article log-length and ChatGPT score: 0.451), Veterinary (0.331), Medicine (0.363), Energy (0.434), and Agricultural and Biological Sciences (0.278). For Veterinary, Medicine, and Agricultural and Biological Sciences, journals from countries where English is not the spoken language were overrepresented in the set of journals with the lowest scores. In some of these, the articles were primarily not in English so the abstracts and titles were translated or constructed separately, either of which may have had a shortening effect. These journals also tended to publish less cited work and some of their articles were summaries of a topic for a local audience rather than primary research (e.g., "Dental diseases in pet rabbits - anatomy, etiology, clinical presentation, differential diagnoses and diagnostic imaging" in *Kleintierpraxis*, a journal "For veterinarians in clinics and practices"). These seemed to have shorter abstracts that described the topic but did not need to mention methods. In Environmental Science, a set of apparently prestigious journals (e.g., high journal impact factors) had long abstracts and high average ChatGPT scores, suggesting that the ChatGPT score association with abstract length might be at least partly a second order effect of journal quality rather than bias. Differences between journals seems to be more important than differences within journals, although the latter is also a factor. This is evident from the fact that correlations between abstract length and ChatGPT score tended to be positive, albeit low: sample size weighted averages of the within-journal correlations for 2023 are Environmental Science (0.169), Veterinary (0.248), Medicine (0.195), Energy (0.108), and Agricultural and Biological Sciences (0.131).
- **Short form contributions with shorter abstracts are included and tend to get lower scores**: This was evident for Veterinary, where some journals had Brief Reports and Case Reports in addition to Standard/Research articles. Case Reports were also in the Medicine dataset. It seems reasonable to expect that short form articles tend to be lower quality than standard articles, so this does not suggest a ChatGPT length bias.

Overall, the association with abstract length could be primarily due to non-bias second order associations, at least from the perspective of REF quality definitions. This assumes that higher quality articles are less likely to be published in national journals and are less likely to be summaries for a national audience. If there is also a bias in terms of ChatGPT favouring longer abstracts, then the size of the bias is less than that suggested by the $R^2$ valued above.

RQ4: International differences in ChatGPT scores have also not been investigated before although international differences in citation counts are understood and used to inform or monitor science policy (Gov.UK, 2022). The use of citations for international comparisons is problematic, however, because of a tendency for national self-citations and the extent to which researchers publish in a national literature that is not indexed in major citation databases (Adams, 1998). ChatGPT does not have the same problems, and the current results suggest that English-speaking countries may not have an advantage, but the evidence is thin and there may be other sources of bias, such as towards research that helps richer countries.

RQ5: The positive correlations with citation counts in all fields found above echo the positive correlations between fields normalised citation counts and the article-level expert quality scores of REF2021 (Thelwall et al., 2023). This is consistent with, but does not prove, the hypothesis that ChatGPT scores tend to positively associate with article quality. This association has been found directly for two small samples (Saad et al., 2024; Thelwall, 2024ab) and for an indirect quality measure (Thelwall & Yaghi, 2024), so this adds to a growing body of evidence about the relationship. This study used single ChatGPT scores whereas previous research has found higher correlations (with quality scores) when averaging thirty ChatGPT scores per article (Thelwall & Yaghi, 2024), so the underlying correlations might be double those in Figure 5. The positive Medicine correlation is unexpected given the negative correlation between REF departmental scores and ChatGPT scores for the UK REF Clinical Medicine category previously found (Thelwall & Yaghi, 2024). This may indicate substantially different scopes for the two categories, a departmental selection effect in the previous study that selected articles from a few departments, or a UK anomaly, but probably not an underlying difference between citations and research quality in this field, given the known positive correlation between these two variables (Thelwall et al., 2023).

## Conclusions

The results give strong evidence that, at least for research in monodisciplinary journals, ChatGPT tends to give higher scores to newer research in all fields, although this tendency is never large and is small in most fields. The results also show substantial disciplinary differences. Together these suggest that the same field normalisation approaches used in citation analysis (Waltman & Van Eck, 2012) are also needed for ChatGPT scores: divide the score for any article by the average score for all articles in the same field and year. This will give a ratio that can be fairly compared between fields and years.

The first author country differences found could indicate ChatGPT bias and/or underlying international differences in the quality of research, with the latter being widely believed to occur by policy makers. Further research is needed to identify whether both are contributors and, if so, the relative balance between them within each field. Without this, international comparisons based on ChatGPT scores have a degree of uncertainty, as do current citation-based comparisons.

Finally, the abstract length factor found potentially indicates another ChatGPT bias, such as against articles in journals with stricter abstract length restrictions, but, from the discussion, it seems more likely that weaker articles are more likely to appear in journals that allow short abstracts or to be for shorter contributions, an acceptable second order effect. Again, more research is needed to investigate this and decide whether it would ever be appropriate to normalise ChatGPT scores for abstract length in addition to field and year.

# Declarations

**Funding and/or Conflicts of interests/Competing interests**: The first author is a member of the Distinguished Reviewers Board of this journal.